\documentclass[prl,twocolumn,showpacs,amsmath,amssymb,superscriptaddress]{revtex4-1}

\usepackage{amssymb}
\usepackage{amsfonts}
\usepackage{amsmath}
\usepackage{graphicx}
\usepackage{xcolor}
\usepackage{ulem}

\newcommand{\etad}{\eta_d}

\newcommand{\vecx}{{\bf x}}

\begin{document}

\title{Dielectric response as a source of viscosity in polar liquids}

\author{David S.\ Dean}

\affiliation{Universit\'e de Bordeaux and CNRS, Laboratoire Ondes et
  Mati\`ere d'Aquitaine, UMR 5798, F-33400 Talence, France}

\author{Haim Diamant}

\affiliation{School of Chemistry and Center for Physics and Chemistry
  of Living Systems, Tel Aviv University, 6997801 Tel Aviv, Israel}

\begin{abstract}
Transport coefficients and dielectric relaxation in liquids are often treated as distinct manifestations of molecular dynamics. We show that, in polar liquids, orientational dipolar fluctuations generate  a substantial contribution to the shear viscosity that can be expressed in terms of  dielectric response parameters. Using a Green-Kubo approach formulated in terms of dipolar body-force correlations, we derive an explicit relation linking the viscosity contribution to the static permittivity and the Debye relaxation time. With a single microscopic cutoff length fixed from one temperature, the theory predicts the temperature dependence of the viscosity for water and several alcohols using independently measured dielectric data. The results identify a general mechanism by which slow polarization dynamics generate an additional, and in strongly polar liquids often dominant, contribution to the viscosity, providing a quantitative bridge between dielectric spectroscopy and rheology.
\end{abstract}

\maketitle

The dielectric constant and viscosity of a liquid determine its suitability for a wide range of applications. An important example is ionic liquids used in battery technologies, whose performance depends on a tradeoff between large dielectric constant and small viscosity~\cite{Hayes15}. In  order to maximize the number of charge carriers,  the solvent should be sufficiently polar to reduce the Born energy of single anions and cations. At the same time, to enhance ionic mobility, the viscosity of the liquid should not be too large.  Accordingly, the dielectric response and shear viscosity (simply referred to as viscosity in this paper) of liquids have been extensively measured in experiments and simulations. Usually the two are studied as separate properties. Here we connect these two properties by studying the contribution of dipole interactions, which underlie dielectric response, to the viscous stress accompanying flow.

The response of a dielectric to an applied electric field is characterized by a frequency-dependent complex dielectric function,
 $ \epsilon(\omega) = \epsilon'(\omega) + i\epsilon''(\omega)$, where $\epsilon(\omega=0)=\epsilon_s$ is the static dielectric constant (permittivity) ~\cite{Kaatze13,Kremer02}. The foundational theory of dielectric spectroscopy, due to Debye~\cite{Debye54}, treats fluid molecules as non-interacting, polar or polarizable thermal rotors, driven by an electric field and damped by the viscosity of the surrounding fluid. The resulting dielectric function,
\begin{equation}
     \epsilon(\omega) = \epsilon_0+ \frac{\chi}{1 + i\omega\tau_D},
    \ \ \ \ \epsilon_s = \epsilon_0 + \chi,
\label{Debye}
\end{equation}
contains a single relaxation mode with relaxation time $\tau_D$.  For the frequency ranges commonly used in dielectric spectroscopy, the contribution of quantum fluctuations to the response is negligible, and $\epsilon_0$ is not necessarily equal to the vacuum permittivity. Here $\chi$ is the effective dipole field polarizability and $\epsilon_0=\epsilon(\omega\rightarrow\infty)$ is the high-frequency limit (sometimes denoted $\epsilon_\infty$).  If the molecules carry a permanent dipole, there is an additional term in $\epsilon_s$ which we absorb in $\chi$. In Debye's theory the relaxation time $\tau_D$ is related to the fluid's viscosity $\eta$ by assuming a Stokes-Einstein relation, 
\begin{equation}
    \tau_D = \frac{\eta\pi a^3}{2T},
\label{DSE}
\end{equation}
where $T$ is the temperature in energy units, and $a$ the particle's effective (hydrodynamic) diameter.

Many extensions and refinements were subsequently  developed for the theory of dielectric response. These are covered in several reviews and books, e.g., Refs.~\cite{Kaatze13,Kremer02}. Various modifications of Eq.~(\ref{Debye}), with different functional forms for $\epsilon(\omega)$, have been widely used to fit experimental data. Such ``non-Debye" relaxations are particularly relevant to complex materials such as supercooled liquids and viscoelastic media~\cite{Feldman02}.

For simple liquids, including water, fitting the measured dielectric function sometimes requires more than one Debye-like relaxation~\cite{Kaatze07,Kaatze17,Ronne99},
\begin{equation}
    \epsilon(\omega) =\epsilon_0 +  \frac{\chi_1}{1 + i\omega\tau_{1}} +
    \frac{\chi_2}{1 + i\omega\tau_{2}} + \ldots.\label{debyeplus}
\end{equation}
The additional, faster relaxations are commonly attributed to specific molecular relaxation mechanisms. 

In this paper we consider an overdamped stochastic dynamical theory for dipole field dynamics, with dipole-dipole interactions, to investigate the contribution of thermal van der Waals interactions to viscosity. According to the Lifshitz theory, the  full many-body behavior of van der Waals interactions is solely encoded in the dielectric response function. Here we aim to derive the contribution of these interactions to the viscosity of liquids.

In Debye's picture the viscosity enters as an independent property of the fluid, externally damping the dipole dynamics. Additional friction experienced by a molecule due to the surrounding dielectric medium has been addressed~\cite{Nee70,Bagchi12}. Studies which use Green-Kubo relations to obtain response from correlation functions do it separately for the dielectric function (from dipolar correlations) and the viscosity (from stress correlations)~\cite{Bagchi12}. 

In this study, we propose a different perspective: the viscosity in large part {\it arises} from the dipoles' stochastic dynamics. Viscosity is ultimately a collective result of molecular interactions, and since dipole interactions are long-ranged and attractive, their contribution to  viscosity is expected to be significant. This point was previously recognized in molecular dynamics simulations~\cite{Lee96}.  We derive a Green-Kubo relation that leads to the following formula for the dipolar contribution to the viscosity:
\begin{equation}
\etad
=
\frac{32T}{45a^3}
\int_0^\infty
\frac{d\omega}{\omega^2}
\frac{
[\operatorname{Im}\epsilon(\omega)]^2
}{
|\epsilon(\omega)|^2
}.
\label{Deta_intro}
\end{equation}
Here $a$ is a length scale coming from a high-wavenumber cutoff equal to $2\pi/a$, i.e., $a$  corresponds to a molecular scale as in Debye's theory. Equation~(\ref{Deta_intro}) holds for any dielectric function $\epsilon(\omega)$ of the form of Eq.~(\ref{debyeplus}), including its generalization to an integral over a continuous spectrum. For the simple Debye model, with dielectric function given by Eq. (\ref{Debye}), we find
\begin{equation}
    \etad = \frac{16\pi T \tau_D \chi^2}{45 a^3(\chi + \epsilon_0)(\chi + 2 \epsilon_0)}.
\label{Deta_intro_simple}
\end{equation}
When $\chi$ is large with respect to $\epsilon_0$, the $\chi$-dependent factors cancel, and $\etad$ coincides with the total viscosity $\eta$ in Debye's phenomenological result, Eq.~(\ref{DSE}), up to a prefactor of order 1 (which can be absorbed into the choice of $a$). This is quite remarkable as the two methods approach the problem from completely different physical standpoints. It implies also that in polar liquids (having large $\epsilon_s$ and therefore large $\chi$), the dipolar contribution to the viscosity must be appreciable.

\begin{figure}
    \centering
\includegraphics[width=0.8\linewidth]{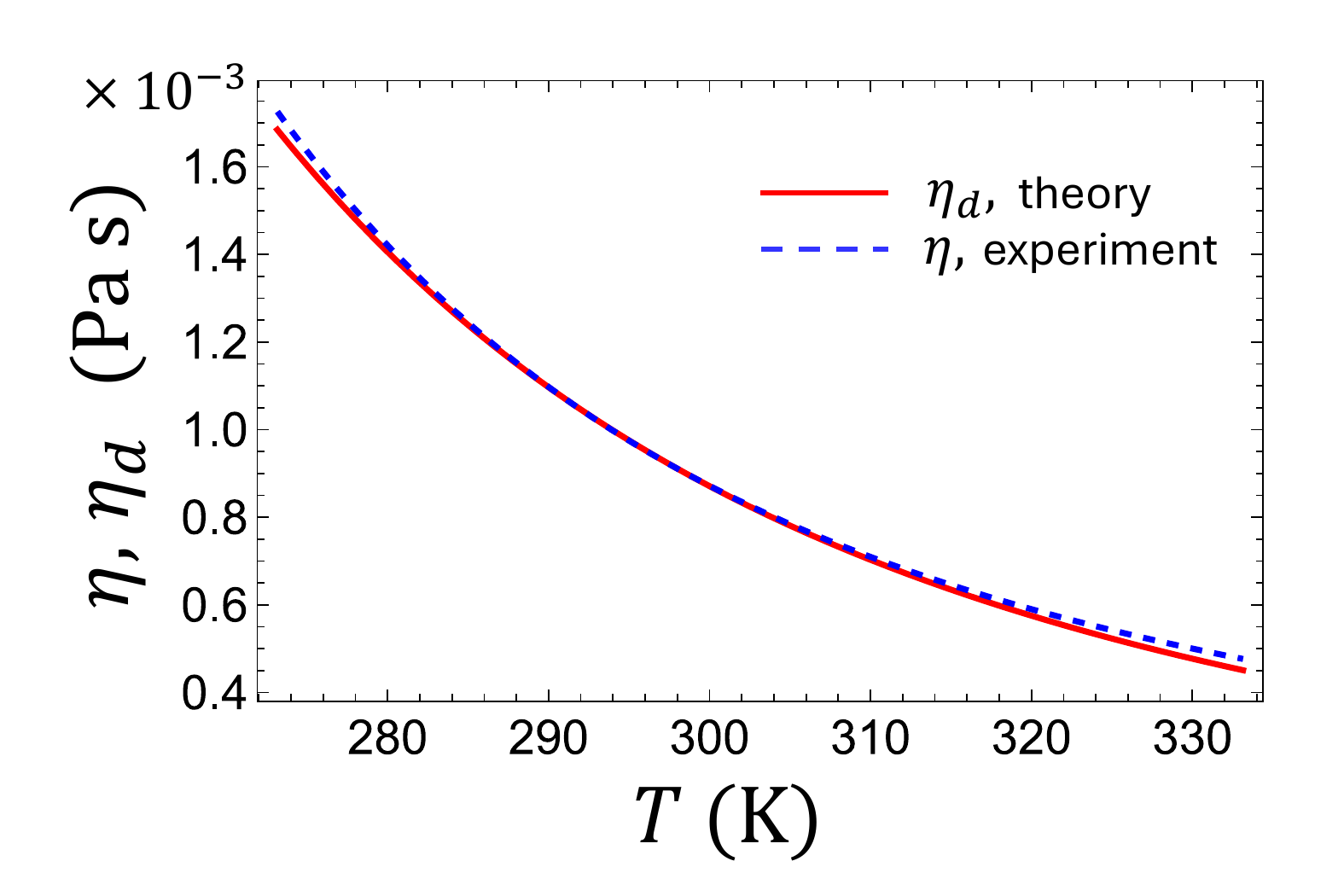}
    \vspace{-0.4cm}
    \caption{Viscosity of water as a function of temperature. The solid line shows the theoretical viscosity contribution obtained from Eq.~(\ref{Deta_intro_simple}) via the procedure described in the text using a single fitted value of $a$. The dashed line is obtained from an empirical interpolation formula for $\eta(T)$ of water~\cite{Viswanath89}.}
    \label{fig:water}
\end{figure}

%[Partly moved to the discussion]
%We will show that the contribution to the viscosity in this model due to thermal van der Waals interactions, solved at the full n-body level, should dominate over other interactions when we compare our predictions to typical viscosities for dielectric liquids. It is intuitively clear that van der Waals interactions being long range and attractive should contribute the viscosity of liquids, and we show that increasing the dielectric constant naturally leads to an increase in viscosity. Our result confirms this intuition. However, we also show that the viscosity thermal van der Waals contribution increases as the typical relaxation time of the local polarisation field increases. If the dipole dynamics becomes extremely fast the contribution to the viscosity goes to zero. For this reason we believe that quantum fluctuations, van der Waals interactions at non-zero Matsubara frequencies will make a much smaller contribution to the viscosity.
%
 \vskip 0.2 truecm
\noindent {\it Stochastic field theory for dipole dynamics.} In this section we  introduce a model for the thermal part of the van der Waals interaction in a dielectric liquid. We will determine the dielectric function $\epsilon(\omega)$ for this model and show how the basic parameters of the model can be determined by fitting experimental dielectric data.

The model is composed of a set of dipole fields, or channels, $p_{\alpha i}({\bf x},t)$, where $\alpha$ is the channel index, and $i=x,y,z$. The Hamiltonian is
\begin{equation}
H = \int d{\bf x} \sum_{\alpha i} \left(\frac{{p}^2_{\alpha i}({\bf x})}{2\chi_\alpha} -\frac{1}{2}{E}_i({\bf x}) {p}_{\alpha i}({\bf x})\right),
\label{H}
\end{equation}
where ${\bf E}$ is the total electric field, given, in terms of the electrostatic potential $\phi$  generated by the dipoles, as
%\begin{equation}
$E_i =-\partial_i \phi$.
%\end{equation}
In the absence of charges, the Poisson equation takes the form
\begin{equation}
\partial_i (\epsilon_0\partial_i \phi) = \partial_i p_i,\label{poisson}
\end{equation}
where $p_i(\vecx,t)=\sum_\alpha p_{\alpha i}(\vecx,t)$ is the total dipole field.
The first term in Eq.~(\ref{H}) is the polarization energy, modeled by a spring-like harmonic self term of each dipole channel $\alpha$ . The parameter $\chi_\alpha$ is the bare polarizability, i.e., the polarizability of the dipole field under a uniform steady applied electric field in the absence of interactions with the other dipoles. We note here that in Eq. (\ref{H}) the self interaction of a dipole with itself is included; however, removing this simply amounts to a renormalization of $\chi_\alpha$.  For the case of a single dipole channel, the Hamiltonian in Eq. (\ref{H}) is the simplest version of a general set of models that can be used to describe dielectric systems \cite{mag06,spr23,bec25}, where additional short-range interactions, of the liquid-crystal type,  can be added to the Hamiltonian. In particular, local gradient terms such as $K\partial_jp_i \partial_jp_i$ may be added to suppress short-wavelength polarization fluctuations and regularize the theory at small length scales. In that more general formulation, the final viscosity expression will retain the same structure, with the cutoff length $a$ replaced by combinations of the corresponding microscopic coefficients. We emphasize, however, that  $a$ represents a molecular-scale structural length and is therefore expected to vary only weakly with temperature compared with the dielectric relaxation time and dielectric constant.

The harmonic model of Eq.~(\ref{H}) has been used to study a number of phenomena in the theory of thermal van der Waals interactions, for instance, the out-of-equilibrium dynamics of the thermal van der Waals force \cite{dea12} and its equilibrium fluctuations~\cite{dea13}. 
The dynamics of the model is taken to be of a stochastic overdamped form given by
\begin{equation}
\frac{\partial p_{\alpha i}}{\partial t} =
-\kappa_\alpha \frac{\delta H}{\delta p_{\alpha i}} = -\kappa_\alpha\left[ \frac{p_{\alpha i}}{\chi_\alpha} + \partial_i\phi \right] +
\sqrt{2\kappa_\alpha T}\eta_{\alpha i},\label{dipd}
\end{equation}
where $\kappa_\alpha $ introduces an intrinsic time scale associated with the dipole dynamics of channel $\alpha$, and $\eta_{\alpha i}({\bf x},t)$ are Gaussian white noise fields of zero mean and with correlation function 
\begin{equation}
    \langle \eta_{\alpha i}({\bf x},t)\eta_{\beta j}({\bf x}',t') \rangle =\delta_{\alpha\beta} \delta_{ij}\delta({\bf x}-{\bf x}') \delta(t-t').\label{cfwn}
\end{equation}

Using this dynamics we now determine the dielectric properties of the model. Fourier transforming Eq.~(\ref{dipd}) in time, we solve for for the dipole fields to find
\begin{equation}
\tilde p_{\alpha i}({\bf x},\omega)= -\chi_\alpha(\omega) \partial_i\tilde \phi({\bf x},\omega ) 
+\chi_\alpha(\omega) \sqrt{\frac{2T}{\kappa_\alpha}}\tilde \eta_{\alpha i}({\bf x},\omega),
\label{ptilde}
\end{equation}
where 
\begin{equation}
\chi_\alpha(\omega) = \frac{\chi_\alpha}{1+i\omega \tau_\alpha}
\label{chiomega}
\end{equation}
is the susceptibility, with $\tau_\alpha = \chi_\alpha/\kappa_{\alpha}$, thus relating $\kappa_\alpha$ to the Debye relaxation time of the dipole channel $\alpha$. 
Substituting Eq.~(\ref{ptilde}) in the Poisson equation (\ref{poisson}) gives
%\begin{eqnarray}
%\epsilon_0 \partial_i^2 \tilde \phi({\bf x},\omega) &=& -\partial_i\sum_\alpha \chi_\alpha(\omega) \partial_i \tilde \phi({\bf x},\omega )+\nonumber \\
%&&\partial_i \sum_{\alpha}\sqrt{\frac{2T}{\kappa_\alpha}}\chi_\alpha(\omega)\tilde \eta_{\alpha i}({\bf x},\omega)
%\end{eqnarray}
%We can now write this as 
\begin{eqnarray}
\partial_i [\epsilon(\omega) \partial_i \tilde \phi({\bf x},\omega)] &=& \partial_i \xi_i({\bf x},\omega),\label{poisson2}\\ 
\epsilon(\omega) &=& \epsilon_0 +\sum_{\alpha} \chi_\alpha(\omega),\label{disp}\\
\xi_i({\bf x},\omega) &=& \sum_{\alpha}\sqrt{2T/\kappa_\alpha}\,\chi_\alpha(\omega)\tilde \eta_{\alpha i}({\bf x},\omega).\label{noise}
\end{eqnarray}
Equations~(\ref{chiomega}) and (\ref{disp}) give the model's dielectric function as a superposition of Debye-like terms, in the form of Eq.~(\ref{debyeplus}), which fits the behavior of a wide range of dielectric materials.

The fluctuating displacement field, generated by the noise $\xi_i$, is found from Eq.~(\ref{poisson2}) as 
\begin{eqnarray}
\tilde D_i({\bf x},\omega) 
%&=& \epsilon_0 \tilde E_i({\bf x},\omega)+ \tilde p%_i({\bf x},\omega)+\xi_i({\bf x},\omega) \nonumber \\&=&
=\epsilon(\omega)\tilde E_i({\bf x},\omega) + \xi_i({\bf x},\omega),\ \ \ \partial_i\tilde D_i=0.
\end{eqnarray}
This microscopic theory turns out to be of the Rytov form for fluctuating electrodynamics \cite{ryt89}. However, we need to derive it here in terms of a dipole field in order to  determine the effect of advection by a nonuniform velocity field, as we now show.

\vskip 0.2cm
 \noindent
 {\it Kubo formula  and viscosity.}  
 The effect of an imposed  flow field $v_i({\bf x})$ is to advect the dipole field, which modifies Eq.~(\ref{dipd}) to (for simplicity, and because we are assuming point-like dipoles, we retain only the advective transport of the polarization field, neglecting rotational \cite{Tsebers80} and stretching couplings to the velocity gradient)
\begin{eqnarray}
  \partial_t p_{\alpha i}({\bf x},t) &+& \nabla_j(v_j({\bf x}) p_{\alpha i}({\bf x},t)) = -\kappa\frac{ \delta H}{\delta p_{\alpha i}({\bf x})} \nonumber \\
  &+& \sqrt{2\kappa T}\eta_{\alpha i}({\bf x},t).
\label{dynamical}
\end{eqnarray}
The body force for the stochastic dynamical theory is given generally by \cite{kru18}
\begin{equation}
f_i({\bf x}) = -p_{\alpha k}({\bf x}) \nabla_i \frac{\delta H}{\delta p_{\alpha k}({\bf x})}.
\end{equation}
In the SM~\cite{SM} we use the Martin-Siggia-Rose path integral formalism \cite{Martin73} to derive the following useful Kubo formula for the change due to advection, within linear response, of a general field $A$ undergoing overdamped stochastic dynamics (the applicability of this formula is much broader than the dynamics treated here),
\begin{equation}
\delta \langle A({\bf x})\rangle= -\frac{1}{T} \int_{0}^\infty dt  \int d{\bf x}' \langle A({\bf x},0)f_j({\bf x}',t)\rangle_c v_j({\bf x}'),  
\label{RA}
\end{equation}
where on the right hand side above $\langle\cdot\rangle_c$ denotes the equilibrium connected correlation function. Specializing to the body force, we get the force generated by the advecting flow as
\begin{equation}
\langle f_i({\bf x}) \rangle = -\frac{1}{ T} \int_{0}^\infty dt\int  d{\bf x}'\langle
  f_i(\vecx,0) f_j(\vecx',t) \rangle v_{j}({\bf x}').
\label{fdt}
\end{equation}

%If the body force is represented via a %stress tensor $\sigma_{ij}$: $f_i %=\partial_j \sigma_{ij}$ we find that %the average of the body force generated %by the advected flow can be written as 
%\begin{equation}
%\langle f_i({\bf x}) \rangle = -\frac{1}%{2 T} \int_{-\infty}^\infty dt\int  d{\bf x}'\left\langle
  %\partial_k \sigma_{ik}(\vecx,0) %\partial'_l \sigma_{jl}(\vecx',t) %\right\rangle_c v_{j}({\bf x}').
%\end{equation}

To extract the change in viscosity, thanks to isotropy, we can consider just the effect of a $z$ dependent flow in the $x$ directon ${\bf v} = v(z) {\bf e}_x$. Representing the force via a stress tensor, $f_i=\partial_j \sigma_{ij}$, and using isotropy and translation invariance, we find that this flow creates an average body force only in the $x$ direction, given by
\begin{equation}
\langle f_x(z)\rangle= \partial_z\int dz' \Lambda(z-z') \partial'_zv_{j}(z'),
\end{equation}
with the viscosity operator
\begin{equation}
\Lambda(z-z')= \frac{1}{T}  \int_{0}^\infty dt \int dx dy \left\langle
   \sigma_{xz}(\vecx-{\bf x}',0)  \sigma_{xz}({\bf 0},t)\right\rangle_c.
\end{equation}
Expansion of the operator $\Lambda$ then gives 
\begin{equation}
\langle f_x(z)\rangle=\etad \partial_z^2\, v(z) + \text{higher derivatives},
\label{expansion}
\end{equation}
with 
\begin{equation}
\etad =  \int_{0}^\infty dt \int d{\bf y} \left\langle
   \sigma_{xz}({\bf y},0)  \sigma_{xz}({\bf 0},t)\right\rangle_c,
\end{equation}
which has the classical Kubo form for the viscosity \cite{kub12}.

Returning to the stochastic dipole theory, we show in \cite{SM} that the stress tensor is given by
\begin{equation}
\sigma_{ij} = \epsilon_0[E_i E_j -\frac{1}{2}\delta_{ij}E^2] + p_j E_i-\sum_\alpha \left(\frac{p^2_{\alpha k}}{2\chi_\alpha}\delta_{ij}\right),
\end{equation} 
leading to
\begin{equation}
\etad =  \int_{0}^\infty dt \int d{\bf y} \left\langle
   E_x({\bf y},0)D_z({\bf y},0)  E_x({\bf 0},t)D_z({\bf 0},t)\right\rangle_c.\label{coreta}
\end{equation}
Due to the Gaussian nature of the theory, the connected correlation function above can be calculated using Wick's theorem and the second-order correlation functions for the fields ${\bf E}$ and ${\bf D}$ \cite{SM}. The latter correlation functions coincide with those obtained from the Rytov theory \cite{ryt89}. This calculation  yields our main result, Eq. (\ref{Deta_intro}).

\noindent {\it Comparison with viscosity data.}
We expect $\etad$ to approach the total viscosity $\eta$ for strongly polar liquids with large $\epsilon_s$. Figure~\ref{fig:water} shows results for water, for which $\epsilon_s(T=298\, \mathrm{K})=78.4\, \epsilon_0$. 
Using measured $\epsilon_s(T)$~\cite{Kaatze07} and $\tau_D(T)$~\cite{Kaatze89}, the only unknown parameter in Eq.~(\ref{Deta_intro_simple}) is the cutoff length $a$. 
Fixing $a$ at a single temperature ($T=298\,\mathrm{K}$) gives $a=3.4$\,\AA, 
consistent with molecular dimensions and the molecular volume of water, $v=(3.1$\,\AA$)^3$, which varies very weakly with temperature. 
All remaining values of $\etad(T)$ then follow without further fitting. The resulting curve agrees very well with the measured viscosity~\cite{Viswanath89}, suggesting that dipole interactions provide the dominant contribution to the viscosity of water. A similar fit can be obtained by fitting $a$ in Debye's theory, Eqs.~(\ref{Debye}) and (\ref{DSE}), which is another indication that, for water, $\eta\simeq\etad$. If the dielectric function is known in more detail than its single-Debye-term representation, one can use the more general relation, Eq.~(\ref{Deta_intro}), to calculate $\eta_d$ more accurately.

\begin{figure}
\centering

\begin{minipage}{0.5\linewidth}
\centering
\includegraphics[width=\linewidth]{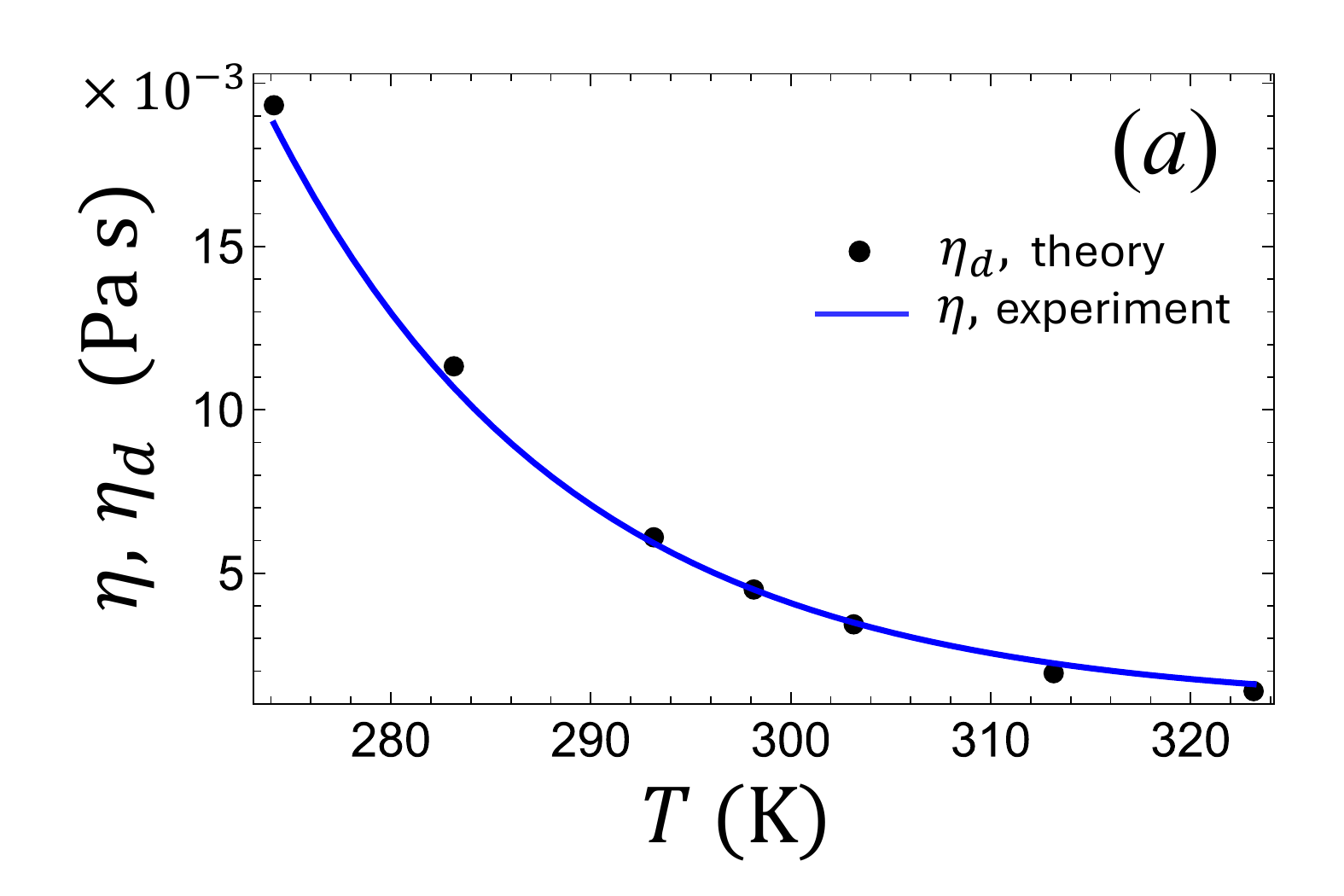}
\end{minipage}%
\hspace{-0.01\linewidth}%
\begin{minipage}{0.5\linewidth}
\centering
\includegraphics[width=\linewidth]{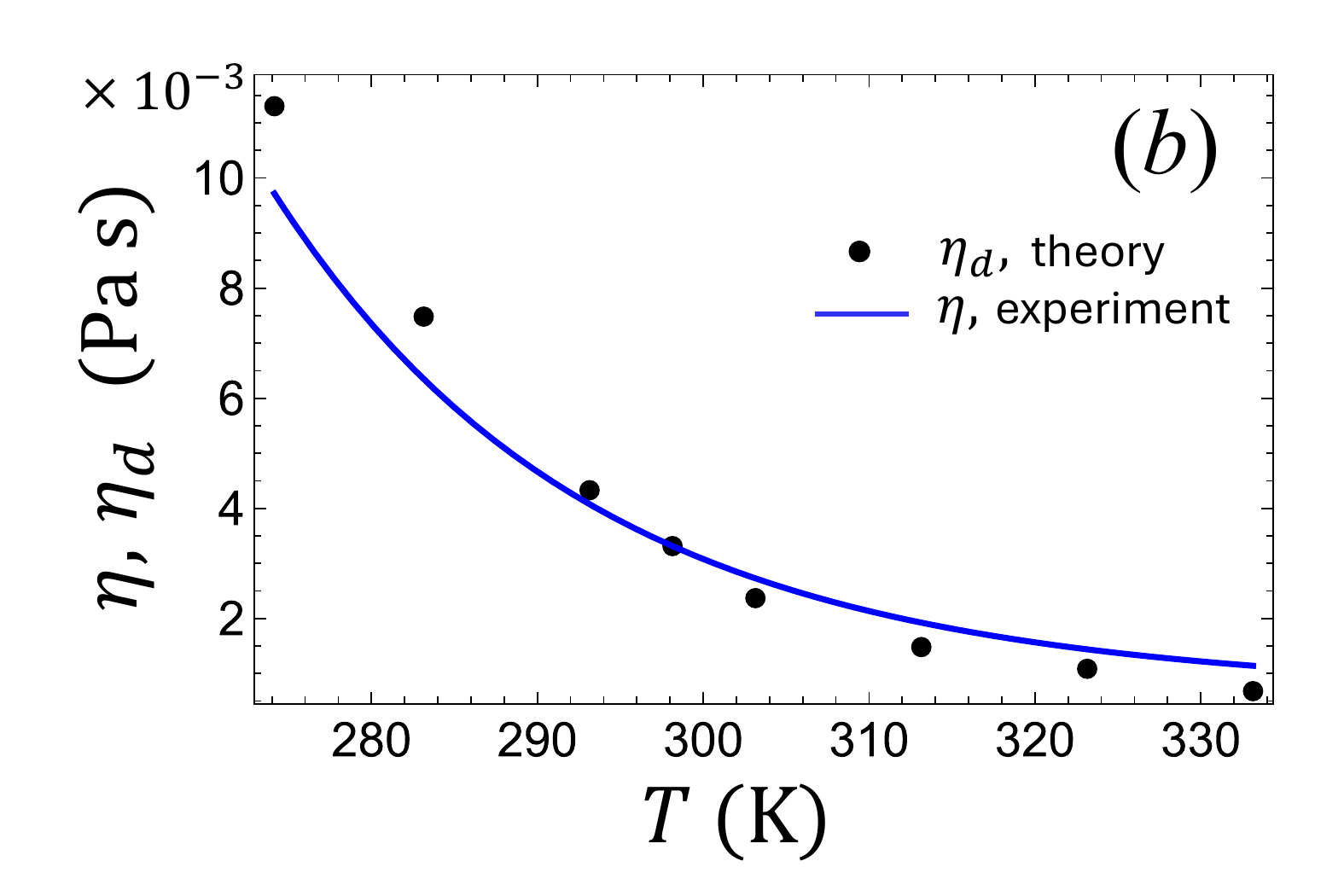}
\end{minipage}

\vspace{0.05cm}

\begin{minipage}{0.5\linewidth}
\centering
\includegraphics[width=\linewidth]{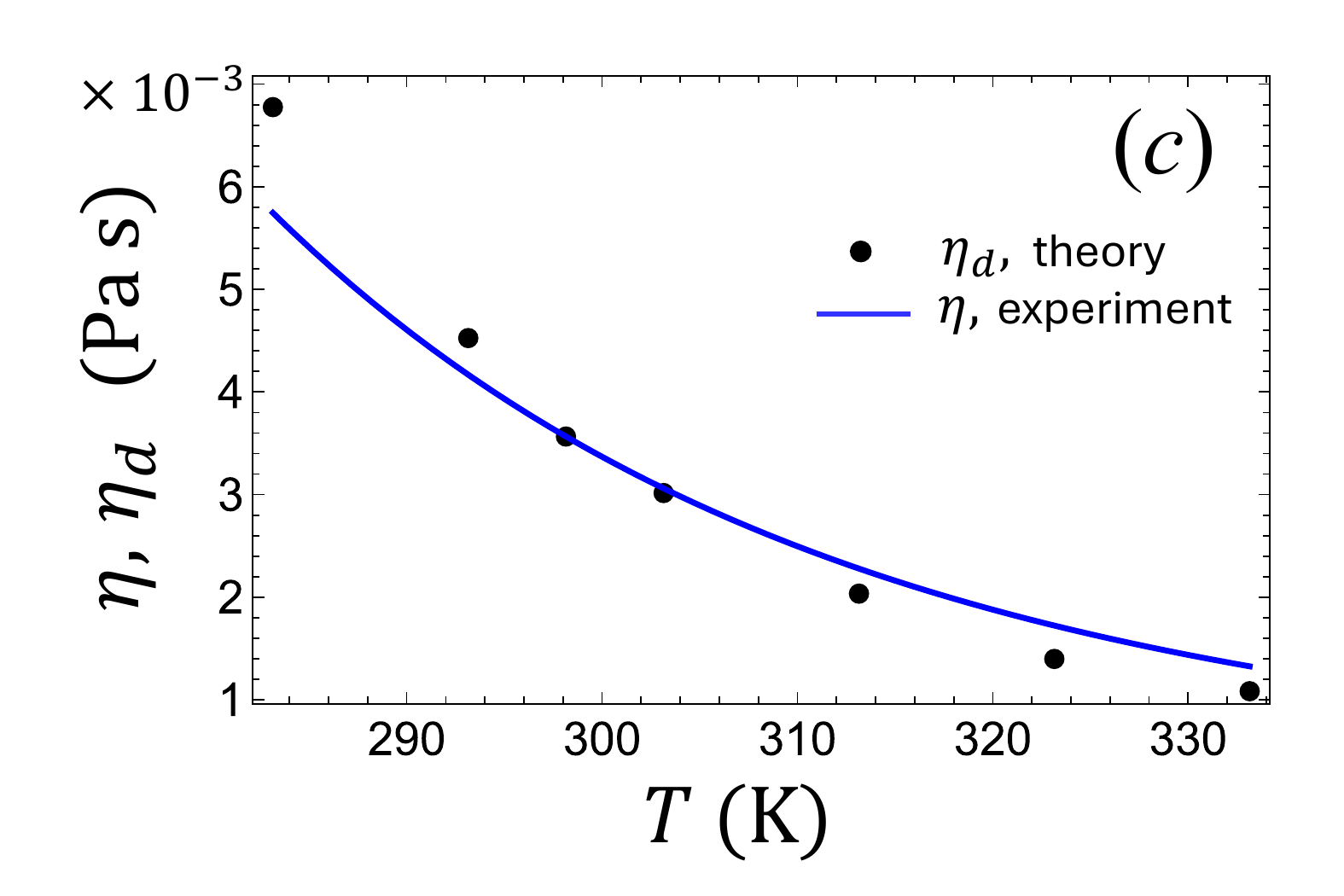}
\end{minipage}%
\hspace{-0.01\linewidth}%
\begin{minipage}{0.5\linewidth}
\centering
\includegraphics[width=\linewidth]{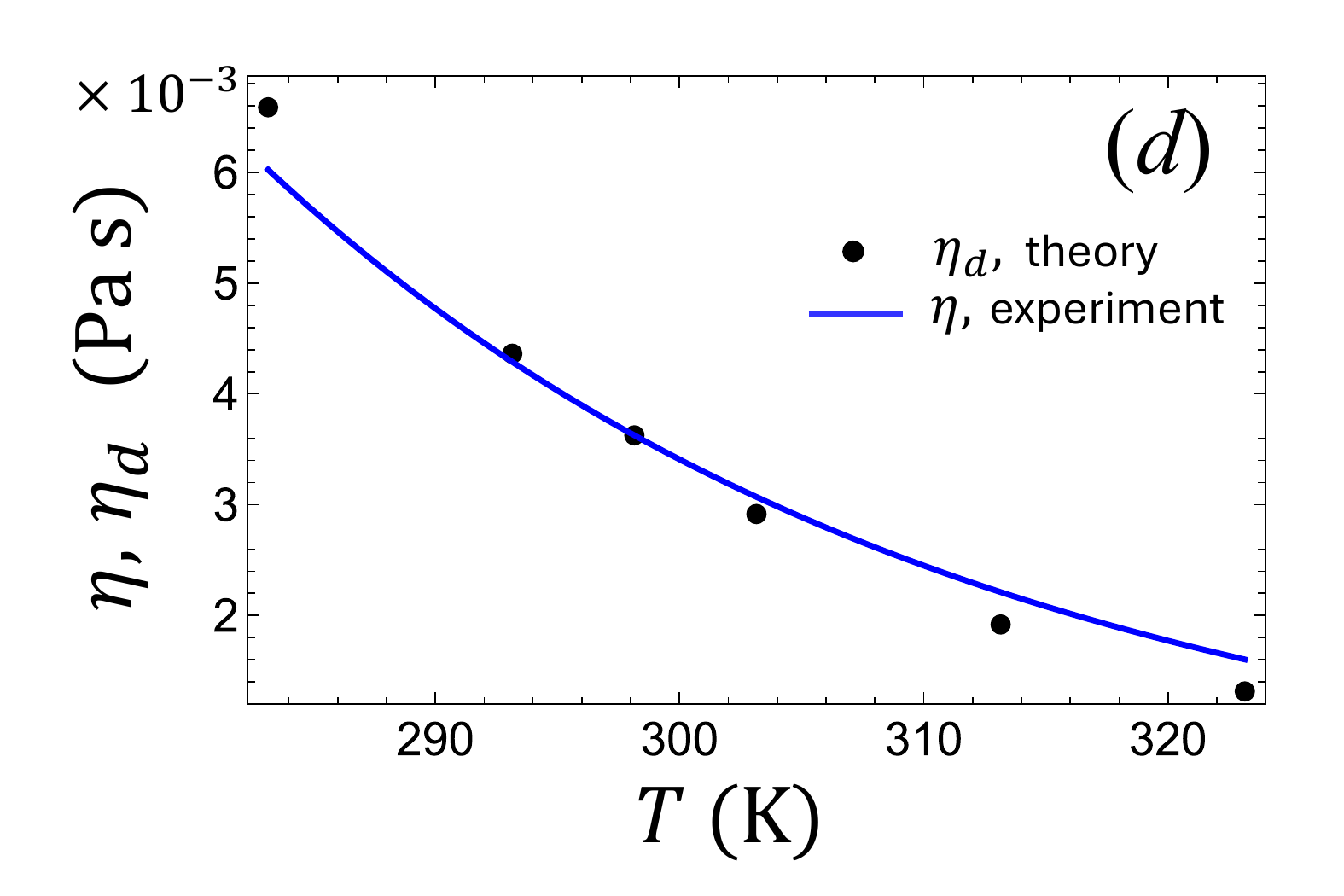}
\end{minipage}

\vspace{0.05cm}

\begin{minipage}{0.5\linewidth}
\centering
\includegraphics[width=\linewidth]{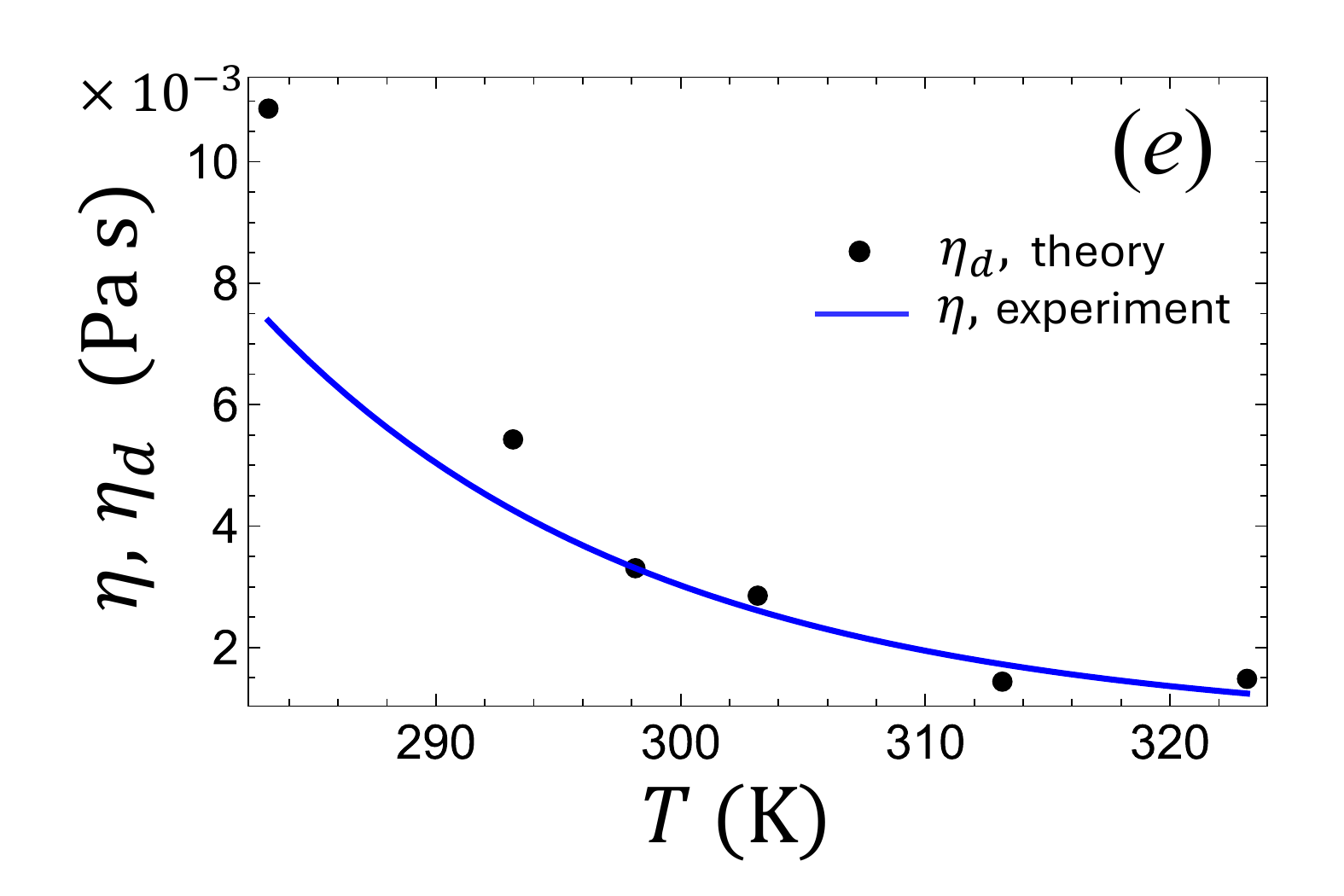}
\end{minipage}

\caption{
Viscosity versus temperature for pentanol isomers:
(a) 3-pentanol,
(b) 2-pentanol,
(c) 1-pentanol,
(d) isopentyl alcohol,
(e) tert-pentanol.
Solid lines: empirical viscosity formulas.
Dots: Eq.~(\ref{Deta_intro_simple}) using data from Ref.~\cite{Kaatze10} with a single fitted value of $a$.
}\label{fig:pentanol}

\end{figure}

Figure~\ref{fig:pentanol} shows the same analysis for five pentanol isomers using the data of Ref.~\cite{Kaatze10}. Their static permittivities at $T=298\,\mathrm{K}$ are $(13.3,13.8,15.1,15.1,5.7)\epsilon_0$ for 3-pentanol, 2-pentanol, 1-pentanol, isopentyl alcohol, and tert-pentanol, respectively. With $a$ fixed at $T=298\,\mathrm{K}$, Eq.~(\ref{Deta_intro_simple}) gives good agreement for 3-pentanol and reasonable agreement for the remaining isomers. The fitted values are $a=8.1,9.0,8.9,9.3,$ and $5.3$\ \AA, respectively, compared with the molecular  volume $v=(5.6$\,\AA$)^3$. Thus one cannot just pick the value of $a$ from the molecular volume and obtain $\eta$ without any fitting. The much larger values for the first four isomers are consistent with the known formation of $\sim 1\,\mathrm{nm}$ dipolar structures that affect Debye relaxation~\cite{Bohmer14,Sillren14}, a property that the more compact tert-pentanol does not possess.
%gives $a=5.3$\,\AA. 
%For 1-propanol, which forms similar $\sim 8$\,\AA  structures~\cite{Sillren14}, the fit gives $a=8.1$\,\AA. These results suggest that the theory captures molecular-scale organization and that alcohol viscosities are likewise largely dipolar in origin.

Finally, a weakly polar (non-hydrogen-bonding) liquid is chlorobenzene (C$_6$H$_5$Cl), for which $\epsilon_s(298\,\mathrm{K})=5.66\,\epsilon_0$, $\tau_D=13.81\,\mathrm{ps}$, and $\eta=0.75\,\mathrm{mPa\cdot s}$~\cite{Pawar02}. Fitting Eq.~(\ref{Deta_intro_simple}) gives $a=3.7\,\mathrm{\AA}$, compared with $v=(5.5\,\mathrm{\AA})^3$, suggesting a dipolar contribution $\etad$ of about 30\% of the total viscosity. An example of a nonpolar but highly polarizable liquid is carbon disulfide (CS$_2$), for which $\epsilon_s(298\,\mathrm{K})=2.62\,\epsilon_0$, $\tau_D=0.63\,\mathrm{ps}$, and $\eta=0.36\,\mathrm{mPa\cdot s}$~\cite{Yu03}. Fitting gives $a=1.3\,\mathrm{\AA}$, significantly smaller than the size inferred from $v=(4.6\,\mathrm{\AA})^3$, implying a small dipolar contribution (a few percent) to the total viscosity. Less polarizable liquids typically show weak frequency dependence and no clear Debye relaxation. This behavior is as expected: the theory is most successful for strongly polar liquids, while for weakly polar systems it predicts only a small dipolar contribution $\etad$ and is not expected to reproduce the full viscosity.

\noindent {\it Conclusions.} Using a  stochastic field theory for dipolar dynamics, we  have shown that the viscosity contribution due to thermal dipolar fluctuation-induced interactions increases with the typical relaxation time of the local polarization field. If the dipole dynamics becomes extremely fast, the contribution to the viscosity becomes negligible. We expect, therefore, that quantum fluctuations, van der Waals interactions at non-zero Matsubara frequencies, not accounted for by the theory, should make a substantially smaller contribution to the viscosity. These trends are physically consistent: the dipolar contribution is largest for strongly polar liquids and becomes subdominant in weakly polar liquids or in systems with short dielectric relaxation times. 

Short-range interactions such as hydrogen bonding between molecules of polar liquids may strongly affect their dipolar response and dynamics~\cite{Zong16,Cardona18}. The model parameters $\chi$ and $\kappa$, therefore, are not properties of an isolated molecule but incorporate short-range effects of the local environment. The values extracted from dielectric measurements correspond to these effective parameters. Obviously, liquids can have dipolar components which are more complex than the terms included in the proposed model. However, it seems that this simple model captures the basic physical mechanisms determining the dielectric properties and viscosity of a wide class of liquids, leading to analytic and testable  predictions. The dielectric function, obtained as a combination of multiple Debye terms [Eq.~(\ref{chiomega}) and (\ref{disp})], and its continuous integral version, can fit a wide range of measured functions. At the same time, even though the multiple Debye terms are additive, each channel adding a factor of $\chi_\alpha(\omega)$, Eq.~(\ref{Deta_intro}) shows that the effect on viscosity is not additive, due to the interaction between dipoles. 

When comparing the theory with experiments, we have focused on the temperature dependence of the viscosity. Since we consider equilibrium liquids of very low compressibility, the same formula is expected to relate the pressure dependence of the viscosity to that of the dielectric properties.

Regarding perspectives, one can readily extend the calculation to higher-order velocity derivatives beyond the second-order Stokes form [see Eq.~(\ref{expansion})] and obtain the corresponding transport coefficients. An intriguing extension would be to study confined liquids \cite{bb94,bb13,car25}, where the effect of dielectric and conducting boundaries on the dipole-induced viscosity near  the confining surfaces could be examined. The interplay between ions and dipoles in electrolytes \cite{Abrashkin07,Adar18} suggests another important direction. Combining stochastic density functional theory for the ions \cite{dea96,kaw94,rob24,bud25} with the field theory used here for the solvent's dipole field may be potentially useful for battery technology. Furthermore,  solvents in batteries are often mixtures, containing a highly polar, viscous liquid,  mixed with a less viscous liquid to avoid an overall high viscosity \cite{wan16}. It would be important to develop a stochastic field theory for dipolar mixtures and their interaction with ionic solutes. 
\vspace{2pt}
\begin{acknowledgments}
\noindent{\it Acknowledgments:} We thank the Institute of Physics, Chinese Academy of Sciences, for its hospitality. H.D.\ is grateful for the hospitality of LOMA, the University of Bordeaux, where this work was initiated. We thank Moshe Kol for helpful input. D.S.D.\ acknowledges support from the grant No. ANR-23-CE30-0020 EDIPS, and from the European Union through the European Research Council by the EMet-Brown (ERC-CoG-101039103) grant. H.D.\ acknowledges support from a joint grant of the Israel Science Foundation
and the National Natural Science Foundation of China (ISF-NSFC Grant No.\ 3159/23) and from the Israel Science Foundation (ISF Grant No.\ 1611/24). 
\end{acknowledgments}

\clearpage
\onecolumngrid

\setcounter{section}{0}

\centerline{\large \bf Supplementary Material}
\vskip 1 truecm
 
In this Supplementary Material, we provide  algebraic details for the analytical results given in the main text. In Sec. (1) we derive the general Kubo relation for the response to advection and its application to the body force of the dipolar field theory [Eqs.~(18) and (19) in the main text]. In Sec.~(2) we calculate the stress tensor of the theory. In Sec.~(3) we apply the Kubo relation to obtain the body force due to a $z$-dependent flow in the $x$ direction [Eq.~(20) of the main text]. Section~(4) presents the derivation of the correlation functions required for calculating the viscosity. Finally, in Sec.~(5) we compute the integral which yields the viscosity. 

\section{1. Kubo relation for an advected field}
\label{sec:Kubo}
Here we consider the dynamics of a general overdamped stochastic field theory for an ensemble of scalar field $\boldmath \phi$ with components $\phi_\alpha$ with general Hamiltonian $H$. The result obtained here will then be applied to the case where the components of $\boldmath \phi$ are the dipole fields $p_{\alpha i}$ of the stochastic dipole model of the main text.

In the absence of any imposed flow the dynamical field equations are given by 
\begin{equation}
\frac{\partial\phi_\alpha({\bf x},t)}{\partial t} = -[M_{\alpha\beta}\frac{\delta H}{\delta \phi_\beta}]({\bf x},t)+ \eta_\alpha({\bf x},t),
\end{equation}
where $M_{\alpha\beta}$ is  the generalized mobility operator and the noise is Gaussian with correlation function
\begin{equation}
\langle \eta_{\alpha}({\bf x},t) \eta_{\beta}({\bf x}',t') \rangle= 2T \delta(t-t')M_{\alpha\beta}({\bf x},{\bf x}') ,
\end{equation}
so the correlation function is white in time, corresponding to Langevin dynamics.
This choice of noise correlation function ensures detailed balance is obeyed and that the equilibrium measure for the fields is the Gibbs-Boltzmann one for the Hamiltonian $H$. This class of models is very general and includes the non-conserved model A case, conserved model B, these models with hydrodynamic interactions and stochastic density functional theories.

When an advecting flow field ${\bf v}$ is imposed, the stochastic dynamics is modified to 
\begin{equation}
\frac{\partial\phi_\alpha({\bf x},t)}{\partial t} +\partial_i[ v_i({\bf x})\phi_\alpha({\bf x},t)]= -[M_{\alpha\beta}\frac{\delta H}{\delta \phi_\beta}]({\bf x},t)+ \eta_\alpha({\bf x},t).\label{dphi}
\end{equation}
The Martin-Siggia-Rose  path integral for the dynamics is  \cite{S-Martin73}
\begin{equation}
Z = \langle\int d[\lambda_\alpha]d[\phi_\alpha] \exp\left(i\int d{\bf x} dt \ \lambda_{\alpha}({\bf x},t)\left[\frac{\partial\phi_\alpha({\bf x},t)}{\partial t} +\partial_i [v_i({\bf x})\phi_\alpha({\bf x},t)] +[M_{\alpha\beta}\frac{\delta H}{\delta \phi_\beta}]({\bf x},t)-\eta_\alpha({\bf x},t)\right]\right)\rangle,
\end{equation}
where $\langle\cdot\rangle$ indicates the average over the white noise. Taking  the average over noise then gives, up to constant prefactors,
\begin{eqnarray}
Z=&&\int d[\lambda_\alpha]d[\phi_\alpha] \exp\left(i\int \ d{\bf x} dt \lambda_{\alpha}({\bf x},t)\left[\frac{\partial\phi_\alpha({\bf x},t)}{\partial t} +\partial_i [v_i({\bf x})\phi_\alpha({\bf x},t)] +[M_{\alpha\beta}\frac{\delta H}{\delta \phi_\beta}]({\bf x},t)\right]\right.\nonumber \\&-&T\left.\int \ dt  d{\bf x}d{\bf x}'
\lambda_\alpha({\bf x},t) M_{\alpha\beta}({\bf x},{\bf x}') \lambda_\beta({\bf x}',t)\right).
\end{eqnarray}
Now integrating over the fields $\lambda_\alpha$ leads to the Onsager-Machlup path integral (again up to constant prefactors),
\begin{eqnarray}
&&Z= \int d[\phi_\alpha]\exp\left(-\frac{1}{4 T} \int d{\bf x} d{\bf x}' dt\right.\nonumber \\&& \left.[\frac{\partial\phi_\alpha({\bf x},t)}{\partial t} +\partial_i [v_i({\bf x})\phi_\alpha({\bf x},t)] +[M_{\alpha\beta}\mu_\beta]({\bf x},t)]M^{-1}_{\alpha\alpha'}({\bf x},{\bf x}') [\frac{\partial\phi_{\alpha'}({\bf x}',t)}{\partial t} +\partial_i [v_i({\bf x})\phi_{\alpha'}({\bf x}',t)] +[M_{\alpha'\beta'}\mu_{\beta'}]({\bf x}',t)]\right)\nonumber \label{MSR},\\
\end{eqnarray}
where we have introduced the chemical potentials for the fields $\phi_\alpha$,
\begin{equation}
\mu_{\alpha}({\bf x}) = \frac{\delta H}{\delta \phi_\alpha({\bf x})}.
\end{equation}
We therefore find that $Z$ is given by 
\begin{equation}
Z = \int d[\phi_\alpha] \exp(-S[\phi,v]),
\end{equation}
where $S$ can be read off from Eq. (\ref{MSR}).
Within this formalism an observable $A({\bf x},0)$ (using time translational invariance we can measure all averages at $t=0$) has  the average 
\begin{equation}
\langle A({\bf x},0)\rangle = \frac{\int d[\phi_\alpha] A({\bf x},0)\exp(-S[\phi,v])}{\int d[\phi_\alpha] \exp(-S[\phi,v])}.
\end{equation}
Now denote by $\delta \langle  A\rangle $ and $\delta S$ the perturbations due to a small applied advecting velocity field to get
\begin{equation}
\delta\langle A({\bf x},0)\rangle = -\frac{\int d[\phi_\alpha] \delta S A({\bf x},0)\exp(-S[\phi,0])}{Z}  + \frac{\int d[\phi_\alpha] A({\bf x},0)\exp(-S[\phi,0])\int d[\phi_\alpha] \delta S \exp(-S[\phi,0])}{(\int d[\phi_\alpha] \exp(-S[\phi,0]))^2},
\end{equation}
which can be written as 
\begin{equation}
\delta\langle A({\bf x},0)\rangle = -\langle \delta S A({\bf x},0)\rangle + \langle \delta S \rangle \langle A({\bf x},0)\rangle= -\langle \delta SA({\bf x},0)\rangle_c,
\label{gk1}
\end{equation}
with $\langle ab\rangle_c = \langle ab\rangle - \langle a\rangle \langle
b\rangle$ and where all averages above are taken with respect to the system when $v_i=0$, that is to say, in equilibrium. 
It is now easy to see that
\begin{equation}
\delta S = 2\times\frac{1}{4T}\int \ d{\bf x}d{\bf x}' dt\  \partial_i [v_i({\bf x})\phi_\alpha({\bf x},t)]M^{-1}_{\alpha\alpha'}({\bf x},{\bf x}')[\frac{\partial\phi_{\alpha'}({\bf x}',t)}{\partial t} +[M_{\alpha'\beta'}\mu_{\beta'}]({\bf x}',t)].
\end{equation}
As the averages in Eq. (\ref{gk1}) are equilibrium averages, the terms with a single derivative with respect to $t$ vanish, and
this gives
\begin{equation}
\delta\langle A({\bf x},0)\rangle = -\langle \delta S' A({\bf x},0)\rangle_c,
\end{equation}
the average taken in the unperturbed equilibrium state, and with
\begin{equation}
\delta S' = 2\times\frac{1}{4T}\int \ d{\bf x}d{\bf x}' dt \ \partial_i [v_i({\bf x})\phi_\alpha({\bf x},t)]M^{-1}_{\alpha\alpha'}({\bf x},{\bf x}') [M_{\alpha'\beta'}\mu_{\beta'}]({\bf x}',t)= \frac{1}{2T} \int \ d{\bf x}dt \ \partial_i [v_i({\bf x})\phi_\alpha({\bf x},t)]\mu_{\alpha}({\bf x},t),
\end{equation}
so the explicit dependence on the operator $M_{\alpha\beta}$ vanishes. This then gives the general Kubo relation
\begin{equation}
\delta \langle A({\bf x},0)\rangle = -\langle \delta S' A({\bf x},0)\rangle_c= -\langle A({\bf x},0)\frac{1}{2T} \int \ d{\bf x}'dt\  \partial_i [v_i({\bf x}')\phi_\alpha({\bf x}',t)]\mu_{\alpha}({\bf x}',t)\rangle_c.
\end{equation}
Now integrating by parts gives
\begin{equation}
\delta \langle A({\bf x},0)\rangle= \frac{1}{2T}\langle A({\bf x},0) \int d{\bf x}'dt \ v_i({\bf x}')\phi_\alpha({\bf x}',t)\partial_i \mu_{\alpha}({\bf x}',t)\rangle_c.
\end{equation}
The body force is given by \cite{S-kru18}
\begin{equation}
f_i({\bf x},t) = -\phi_\alpha({\bf x})\partial_i \mu_{\alpha}({\bf x},t),
\end{equation}
(with summation implied over the index $\alpha$), having a contribution from each field component $\alpha$. 
Thus we find the general Kubo formula:
\begin{equation}
\delta \langle A({\bf x},0)\rangle= -\frac{1}{2T}\langle A({\bf x},0) \int d{\bf x}'dt \ v_j({\bf x}')f_j({\bf x}',t)\rangle_c. \label{RAmain1}
\end{equation}
In the above the time integral was over $(-\infty, \infty)$, however, as the connected correlation function is the equilibrium one we, can use the equilibrium identity $\langle A({\bf x},0)f_i({\bf x}',t)\rangle_c= \langle A({\bf x},0)f_i({\bf x}',-t)\rangle_c$ and write the integral over $(0,\infty)$, as is the convention in Kubo formulas, to get :
\begin{equation}
\delta \langle A({\bf x},0)\rangle=-\frac{1}{T}\int_0^\infty dt  \int d{\bf x}' \ \langle A({\bf x},0)f_j({\bf x}',t)\rangle_c\ v_j({\bf x}'), \label{RAmain2}
\end{equation}
which is Eq. (18) of the main text.

Now if we choose $A({\bf x},0)= f_i({\bf x},0)$ we find 
\begin{equation}
\delta \langle f_i({\bf x},0)\rangle= -\frac{1}{T} \int_0^\infty dt \int d{\bf x}'\ \langle f_i({\bf x},0)f_j({\bf x}',t)\rangle_c v_j({\bf x}'),
\label{deltaf}
\end{equation}
which is equivalent to Eq.~(19) in the main text,  because in equilibrium $\langle f_i({\bf x},t)\rangle =0$, {\it i.e.}, the advective perturbation is $\langle\delta f_i({\bf x},t)\rangle=\langle f_i({\bf x},t)\rangle$. 
Alternatively we can write ,
\begin{equation}
 \langle \frac{\delta f_i({\bf x},0)}{\delta v_j({\bf x}')}\rangle= -
 \frac{1}{T}\int_0^\infty dt\langle  \ f_i({\bf x},0)f_j({\bf x}',t)\rangle_c.\label{f}
\end{equation}
We emphasize that Eq. (\ref{f}) is  valid for the stochastic dynamics of the fields $p_{\alpha i}({\bf x})$ considered in the main text, but in fact holds for any dynamics of the form in Eq. (\ref{dphi}) and any Hamiltonian $H$ (that is to say not only for quadratic Hamiltonians).

\section{2. The stress tensor for the dipole field theory}
\label{sec:stress}

Here we derive the stress tensor $\sigma_{ij}$ such that the body force is given by $f_i=\partial_j \sigma_{ij}$, starting from the body force 
\cite{S-kru18}
\begin{equation}
f_i({\bf x},t) = -p_{\alpha k}({\bf x},t)\partial_i \frac{\delta H}{\delta p_{\alpha k}({\bf x})}.
\end{equation}
For the dipole field model we obtain 
\begin{equation}
f_i({\bf x},t)= -\sum_{\alpha,i}p_{\alpha j}({\bf x},t)\partial_i \left(\frac{p_{\alpha j}({\bf x},t)}{\chi_\alpha} -{E}_j({\bf x},t)\right)
=-\partial_i\sum_\alpha \left(\frac{p^2_{\alpha k}({\bf x},t)}{2\chi_\alpha}\delta_{ij}\right) +p_j({\bf x},t)\partial_iE_j({\bf x},t) ,
\end{equation}
where ${\bf E}({\bf x},t)$ is the electric field, and 
\begin{equation}
p_j({\bf x},t)= \sum_\alpha p_{\alpha j}({\bf x},t)
\end{equation}
is the total dipole field. The charge density is given by $\rho=-\partial_jp_j$. In terms of the electrostatic field one has, via the standard computation for the stress tensor,
\begin{equation}
\rho E_i= \epsilon_0 E_i  \partial_j E_j = \epsilon_0 \partial_j[ E_iE_j ] - \epsilon_0E_j  \partial_i E_j = \epsilon_0\partial_j[E_i E_j -\frac{1}{2}\delta_{ij}E^2].
\end{equation}
This gives
\begin{equation}
f_i({\bf x},t)= -\partial_i\sum_\alpha \left(\frac{p^2_{\alpha k}}{2\chi_\alpha}\delta_{ij}\right) + \partial_j (p_j E_i)+ \epsilon_0\partial_j[E_i E_j -\frac{1}{2}\delta_{ij}E^2],
\end{equation}
and so we obtain the stress tensor as
\begin{equation}
\sigma_{ij} = \epsilon_0[E_i E_j -\frac{1}{2}\delta_{ij}E^2] + p_j E_i-\frac{\delta_{ij}}{2}\sum_\alpha \left(\frac{p^2_{\alpha k}}{\chi_\alpha}\right).
\end{equation}

\section{3. Body force created by a $z$-dependent flow in the $x$ direction}
\label{sec:zflow}

We start from Eq.~(19) of the main text [or Eq.~(\ref{deltaf}) above],
\begin{equation}
\langle f_i({\bf x}) \rangle = -\frac{1}{T} \int_{0}^\infty dt\int  d{\bf x}'\left\langle
  \partial_k \sigma_{ik}(\vecx,0) \partial'_l \sigma_{jl}(\vecx',t) \right\rangle_c v_{j}({\bf x}').
\end{equation} 
and consider the case where ${\bf v}({\bf x}) = v(z) {\bf e}_x$. This then gives
\begin{equation}
\langle f_i({\bf x}) \rangle = -\frac{1}{T} \int_{0}^\infty dt\int  d{\bf x}'\left\langle
  \partial_k \sigma_{ik}(\vecx,0) \partial'_l \sigma_{xl}(\vecx',t) \right\rangle_c v(z').
\end{equation}
Now notice that $\langle
  \partial_k \sigma_{ik}(\vecx,0) \partial'_l \sigma_{xl}(\vecx',t) \rangle$ depends on ${\bf x}-{\bf x}'$ by translational invariance and so the derivatives over $k=x$ and $k=y$ above yield zero as $\partial_x=-\partial'_x$ and $\partial_y=-\partial'_y$. We thus have
 \begin{equation}
\langle f_i({\bf x}) \rangle = -\frac{1}{T} \int_{0}^\infty dt\int  d{\bf x}'\left\langle
  \partial_z \sigma_{iz}(\vecx,0) \partial'_l \sigma_{xl}(\vecx',t) \right\rangle_c v(z').
\end{equation} 
Similarly one can integrate the term $\partial'_l \sigma_{xl}(\vecx',t)$ for $l=x$ and $l=y$ to obtain zero, and so
\begin{equation}
\langle f_i({\bf x}) \rangle = -\frac{1}{T} \int_{0}^\infty dt\int  d{\bf x}'\left\langle
  \partial_z \sigma_{iz}(\vecx,0) \partial'_z \sigma_{xz}(\vecx',t) \right\rangle_c v(z').
\end{equation}
Integrating by parts over $z'$ then gives
\begin{equation}
\langle f_i({\bf x}) \rangle = \frac{1}{T} \int_{0}^\infty dt\int  d{\bf x}'\left\langle
  \partial_z \sigma_{iz}(\vecx,0) \sigma_{xz}(\vecx',t) \right\rangle_c \partial'_z v(z').
\end{equation}
Setting $i=x$ gives Eq.~(20) of the main text upon the identification
\begin{equation}
  \int_{0}^\infty dt\int  dx' dy' \left\langle
   \sigma_{xz}(\vecx,0) \sigma_{xz}(\vecx',t) \right\rangle_c= \int_{-\infty}^\infty dt\int  dx' dy' \left\langle
  \sigma_{xz}(\vecx-\vecx',0) \sigma_{xz}({\bf 0},t) \right\rangle_c=\Lambda(z-z').
\end{equation}
Notice that one has a $z$-component of the body force,
\begin{equation}
\langle f_z({\bf x}) \rangle = \frac{1}{T} \int_{0}^\infty dt\int  d{\bf x}'\left\langle
  \partial_z \sigma_{zz}(\vecx,0) \sigma_{xz}(\vecx',t) \right\rangle_c \partial'_z v(z'),
\end{equation}
which is not obviously zero. However, the first-order term in the operator expansion gives
\begin{equation}
\langle f_z({\bf x}) \rangle = \frac{1}{T}\partial_z \int_{-\infty}^\infty dt\int  d{\bf y}\left\langle
   \sigma_{zz}({\bf y},0) \sigma_{xz}({\bf 0},t) \right\rangle_c \partial_z v(z).
\end{equation}
By isotropy the tensor 
\begin{equation}
\Sigma_{ijkl}= \int_{0}^\infty dt\int  d{\bf y}\left\langle
   \sigma_{ij}({\bf y},0) \sigma_{kl}({\bf 0},t) \right\rangle_c = A \delta_{ij}\delta_{kl}+ B \delta_{ik}\delta_{jl}+ C \delta_{il}\delta_{jk},
\end{equation}
which implies $\langle f_z({\bf x}) \rangle=0$.

\section{4. Correlation functions and the link with fluctuating electrodynamics}
\label{sec:correlations}

Here we compute the correlation functions of the  electric field ${\bf E}$ and displacement field ${\bf D}$ arising from the stochastic dipole model and which are necessary to compute the viscosity via Eq.~(25) of the main text. We show that the statistics of the electric field reproduces the Rytov theory of fluctuating electrodynamics \cite{S-ryt89}.

To start with we define the Green's function
\begin{equation}
 \partial_i\epsilon(\omega)\partial_i G({\bf x},{\bf x}',\omega)=-\delta({\bf x}-{\bf x}').
 \end{equation}
We use the above notation because, in other models, $\epsilon(\omega)$ could be also space-dependent. The solution for the electrostatic potential is then given by
 \begin{equation}
  \tilde \phi({\bf x},\omega)= -\int d{\bf x}' G({\bf x},{\bf x}',\omega)\partial'_i  \xi_i({\bf x}',\omega),
 \end{equation}
where $\xi_i({\bf x},\omega)$ is the noise given in Eq.~(14) of the main text.
This gives the electric field as
\begin{equation}
\tilde E_i({\bf x},\omega) = -\partial_i \phi = \int d{\bf x}' \partial_iG({\bf x},{\bf x}',\omega)\partial'_j \xi_i({\bf x}',\omega)=-\int d{\bf x}' [\partial_i\partial_j'G({\bf x},{\bf x}',\omega)] \xi_i({\bf x}',\omega),
\end{equation}
and we recall that the displacement field is given by
\begin{equation}
\tilde D_i({\bf x},\omega)= \epsilon(\omega)\tilde E_i({\bf x},\omega)+ \xi_i({\bf x},\omega).
\label{DeE}
\end{equation}

The noise $\xi_i({\bf x},\omega)$ is Gaussian and has zero mean and the correlation function
\begin{equation}
\langle \xi_i({\bf x},\omega)\xi_j({\bf x}',\omega')\rangle = 4\pi T \delta(\omega+\omega')\delta_{ij} \delta({\bf x}-{\bf x}')\sum_{\alpha}\frac{\chi_\alpha(\omega)\chi_\alpha(-\omega)}{\kappa_\alpha}.
\end{equation}
We notice that
\begin{equation}
\frac{\chi_\alpha(\omega)\chi_\alpha(-\omega)}{\kappa_\alpha}=\frac{\chi_\alpha^2 }{\kappa_\alpha(1 +\omega^2 \tau_\alpha^2)}=\frac{\chi_\alpha\tau_\alpha }{1 +\omega^2 \tau_\alpha^2},
\end{equation}
while
\begin{equation}
{\rm Im}\,\chi_\alpha(\omega) = -\frac{\chi_\alpha \omega\tau_\alpha}{1 +\omega^2 \tau_\alpha^2}.
\end{equation}
This then gives [see Eq.~(13) in the main text]
\begin{equation}
\langle \xi_i({\bf x},\omega)\xi_j({\bf x}',\omega')\rangle=-\frac{4\pi T }{\omega}\delta_{ij}\delta({\bf x}-{\bf x}')\delta(\omega+\omega')\sum_{\alpha}{\rm Im}\,\chi_\alpha(\omega)= -\frac{4\pi T }{\omega}\delta_{ij}\delta({\bf x}-{\bf x}')\delta(\omega+\omega'){\rm Im}\,\epsilon(\omega).
\end{equation}
The electric-field correlation is
\begin{equation}
\left\langle
\tilde E_i(\mathbf{x},\omega)
\tilde E_j(\mathbf{x}',\omega')
\right\rangle
=
-\frac{4\pi T}{\omega}
\delta(\omega+\omega')
\int d\mathbf{x}''
\,{\rm Im}[\epsilon(\omega)]
\left[
\partial_k''\partial_i
G(\mathbf{x},\mathbf{x}'',\omega)
\right]
\left[
\partial_k''\partial_j'
G(\mathbf{x}',\mathbf{x}'',-\omega)
\right].
\end{equation}
Taking the derivatives with respect to the external coordinates outside the
integral gives
\begin{equation}
\left\langle
\tilde E_i(\mathbf{x},\omega)
\tilde E_j(\mathbf{x}',\omega')
\right\rangle
=
-\frac{4\pi T}{\omega}
\delta(\omega+\omega')
\partial_i\partial_j'
\int d\mathbf{x}''
\,{\rm Im }[\epsilon(\omega)]
\partial_k''
G(\mathbf{x},\mathbf{x}'',\omega)
\partial_k''
G(\mathbf{x}',\mathbf{x}'',-\omega).
\end{equation}
Integrating by parts with respect to \(\mathbf{x}''\) gives
\begin{equation}
\left\langle
\tilde E_i(\mathbf{x},\omega)
\tilde E_j(\mathbf{x}',\omega')
\right\rangle
=
\frac{4\pi T}{\omega}
\delta(\omega+\omega')
\partial_i\partial_j'
\int d\mathbf{x}''
\,G(\mathbf{x},\mathbf{x}'',\omega)
\partial_k''
\left[
{\rm Im}[\epsilon(\omega)]
\partial_k ''
G(\mathbf{x}',\mathbf{x}'',-\omega)
\right].
\label{EE}
\end{equation}

Equation (\ref{EE}) appears quite unwieldy. However, a major simplification arises using what is called the {\it resolvent identity}. In operator notation the equation for the Green's function $G$ is  
\begin{equation}
H(\omega)G(\omega)=I,
\end{equation}
where
\begin{equation}
H(\omega)
=
-\partial_i \epsilon(\omega)\partial_i.
\end{equation}
Thus,
\begin{equation}
G(\omega)=H^{-1}(\omega).
\end{equation}
Using
\begin{equation}
A^{-1}(B-A)B^{-1}
=
A^{-1}-B^{-1},
\end{equation}
with \(A=H\) and \(B=H^\dagger\), gives
\begin{equation}
G(H^\dagger-H)G^\dagger
=
G-G^\dagger.
\end{equation}
Since
\begin{equation}
H^\dagger(\omega)
=
-\partial_i
\epsilon^*(\omega)\partial_i,
\end{equation}
we have
\begin{equation}
H^\dagger-H
=
-\partial_i
\bigl(
\epsilon^*(\omega)
-
\epsilon(\omega)
\bigr)\partial_i
=
2i\,\partial_i
{\rm Im}[\epsilon(\omega)]\partial_i
.
\end{equation}
Therefore,
\begin{equation}
G-G^\dagger
=
2i\,
G
\partial_i
{\rm Im}[\epsilon(\omega)]\,\partial_i
G^\dagger,
\end{equation}
which in coordinate representation is
\begin{equation}
G(\mathbf{x},\mathbf{x}',\omega)
-
G^\dagger(\mathbf{x},\mathbf{x}',\omega)
=
2i
\int d\mathbf{x}''
\,G(\mathbf{x},\mathbf{x}'',\omega)
\partial_i''
{\rm Im}[\epsilon(\omega)]
\partial_i''
G^\dagger(\mathbf{x}'',\mathbf{x}',\omega).
\end{equation}
Note that by the definition of the adjoint kernel we have 
\begin{equation}
G^\dagger(\mathbf{x}'',\mathbf{x}',\omega)
=
G^*(\mathbf{x}',\mathbf{x}'',\omega),
\end{equation}
and also 
\begin{equation}
\epsilon(-\omega)=\epsilon^*(\omega),
\end{equation}
so that
\begin{equation}
G(\mathbf{x}',\mathbf{x}'',-\omega)
=
G^*(\mathbf{x}',\mathbf{x}'',\omega)
=
G^\dagger(\mathbf{x}'',\mathbf{x}',\omega).
\end{equation}
The resolvent identity therefore gives
\begin{equation}
2i
\int d\mathbf{x}''
\,G(\mathbf{x},\mathbf{x}'',\omega)
\partial_i''
\left[
{\rm Im}[\epsilon(\omega)]
\partial_i''
G(\mathbf{x}',\mathbf{x}'',-\omega)
\right]
=
G(\mathbf{x},\mathbf{x}',\omega)
-
G^\dagger(\mathbf{x},\mathbf{x}',\omega).
\end{equation}
Now $G$ is self adjoint as a spatial operator
\begin{equation}
G(\mathbf{x},\mathbf{x}',\omega)
=
G(\mathbf{x}',\mathbf{x},\omega),
\end{equation}
and therefore
\begin{equation}
G^\dagger(\mathbf{x},\mathbf{x}',\omega)
=
G^*(\mathbf{x},\mathbf{x}',\omega).
\end{equation}
This now gives
\begin{equation}
G(\mathbf{x},\mathbf{x}',\omega)
-
G^\dagger(\mathbf{x},\mathbf{x}',\omega)
=
2i\,
{\rm Im}(G(\mathbf{x},\mathbf{x}',\omega)),
\end{equation}
and therefore
\begin{equation}
\int d\mathbf{x}''
\,G(\mathbf{x},\mathbf{x}'',\omega)
\partial_i''
\left[
{\rm Im}[\epsilon(\omega)]
\partial_i''
G(\mathbf{x}',\mathbf{x}'',-\omega)
\right]
=
{\rm Im}(G(\mathbf{x},\mathbf{x}',\omega)).
\end{equation}
This finally leads to 
\begin{equation}
\left\langle
\tilde E_i(\mathbf{x},\omega)
\tilde E_j(\mathbf{x}',\omega')
\right\rangle
=
\frac{4\pi T}{\omega}
\delta(\omega+\omega')
\left[
\partial_i\partial_j'
{\rm Im}(G(\mathbf{x},\mathbf{x}',\omega))
\right].
\end{equation}
This is the same correlation function as the thermal or classical high-temperature component of the Rytov theory of fluctuation electrodynamics in the non-retarded limit \cite{S-ryt89}

In what follows it is convenient to define
\begin{equation}
Q_{ij}(\mathbf{x},\mathbf{x}',\omega)
=
\partial_i\partial_j'
G(\mathbf{x},\mathbf{x}',\omega),
\end{equation}
and so 
\begin{equation}
\langle \tilde E_i(\mathbf{x},\omega)\tilde E_j(\mathbf{x}',\omega')\rangle
=
\frac{4\pi T}{\omega}
\delta(\omega+\omega')
{\rm Im}[Q_{ij}(\mathbf{x},\mathbf{x}',\omega)].
\end{equation}

We now compute the other correlation functions necessary to compute the viscosity correction. We find 
\begin{equation}
\langle \tilde E_i(\mathbf{x},\omega)
{\xi}_j(\mathbf{x}',\omega')\rangle
=
\frac{4\pi T}{\omega}
\delta(\omega+\omega')
{\rm Im}[\epsilon(\omega)]
Q_{ij}(\mathbf{x},\mathbf{x}',\omega),
\end{equation}
and 
\begin{equation}
\langle {\xi}_i(\mathbf{x},\omega)
\tilde E_j(\mathbf{x}',\omega')\rangle
=
\frac{4\pi T}{\omega'}
\delta(\omega+\omega')
{\rm Im}[\epsilon(\omega')]
Q_{ji}(\mathbf{x}',\mathbf{x},\omega')= \frac{4\pi T}{\omega}
\delta(\omega+\omega')
{\rm Im}[\epsilon(\omega)]
Q^*_{ij}(\mathbf{x},\mathbf{x}',\omega),
\end{equation}
where we have used ${\rm Im}[\epsilon(-\omega)]= -{\rm Im}[\epsilon(\omega)]$. From these last two results and Eq.~(\ref{DeE}) we then find
\begin{equation}
\langle \tilde E_i(\mathbf{x},\omega)
\tilde D_j(\mathbf{x}',\omega')\rangle
=
\frac{4\pi T}{\omega}
\delta(\omega+\omega')
\Big[
\epsilon^*(\omega)
\operatorname{Im}[Q_{ij}(\mathbf{x},\mathbf{x}',\omega)]
+{\rm Im}[\epsilon(\omega)]
Q_{ij}(\mathbf{x},\mathbf{x}',\omega)
\Big].
\end{equation}
Now using
\begin{equation}
a^*\,\operatorname{Im}b
+(\operatorname{Im}a)b
=
\operatorname{Im}(ab),
\end{equation}
this simplifies to
\begin{equation}
\langle \tilde E_i(\mathbf{x},\omega)
\tilde D_j(\mathbf{x}',\omega')\rangle
=
\frac{4\pi T}{\omega}
\delta(\omega+\omega')
\operatorname{Im}
\left[
\epsilon(\omega)
Q_{ij}(\mathbf{x},\mathbf{x}',\omega)
\right].
\end{equation}

As for the displacement-displacement correlation,
\begin{align}
\langle D_i (\mathbf{x},\omega)D_j(\mathbf{x}',\omega')\rangle
&=
\epsilon(\omega)
\epsilon(\omega')
\langle \tilde E_i(\mathbf{x},\omega)\tilde E_j(\mathbf{x}',\omega')\rangle
+
\epsilon(\omega)
\langle \tilde E_i(\mathbf{x}',\omega')\xi_j(\mathbf{x}',\omega')\rangle
\nonumber\\
&\quad+
\epsilon(-\omega)
\langle {\xi}_i(\mathbf{x},\omega)\tilde E_j(\mathbf{x}',\omega')\rangle
+
\langle {\xi}_i(\mathbf{x},\omega) {\xi}_j(\mathbf{x}',\omega')\rangle.
\end{align}
Substituting the preceding expressions and using the identity
\begin{equation}
aa^* \operatorname{Im}[b] + a \operatorname{Im}[a] b + a^*\operatorname{Im}[a] b^*= \operatorname{Im}[a^2 b],
\end{equation}
then gives
\begin{equation}
\langle \tilde D_i(\mathbf{x},\omega)
\tilde D_j(\mathbf{x}',\omega')\rangle
=
\frac{4\pi T}{\omega}
\delta(\omega+\omega')
\Bigg\{
\operatorname{Im}
\Big[
\epsilon(\omega)
\epsilon(\omega)
Q_{ij}(\mathbf{x},\mathbf{x}',\omega)
\Big]
-
\delta_{ij}
\operatorname{Im}[\epsilon(\omega)]
\delta(\mathbf{x}-\mathbf{x}')
\Bigg\}.
\end{equation}
We now have all the correlation functions necessary to compute the viscosity.

\section{5. Doing the integral to obtain the viscosity}
\label{sec:integral}

Here we show how to compute the integral
\begin{equation}
\mathcal{I}_{ij}({\bf x}-{\bf x}')
= \int_{0}^\infty dt\int  d{\bf x}\left\langle
   \sigma_{ij}({\bf x},0) \sigma_{kl}({\bf x}',t) \right\rangle_c =\frac{1}{2}
\int_{-\infty}^{\infty} dt\,
\left\langle
D_i(\mathbf{x},t) E_j(\mathbf{x},t)
D_i(\mathbf{x}',0) E_j(\mathbf{x}',0)
\right\rangle_{\mathrm{c}},
\qquad i\neq j.
\end{equation}
(taking the integration range over $(-\infty,\infty)$ to simplify the Fourier analysis) appearing  in Eq. (21) of the main text to obtain $\eta_d$.

Since \(D_i\) and \(E_j\) are linear functions of Gaussian noise, they are jointly Gaussian. Wick's theorem therefore gives
\begin{equation}
\left\langle
D_i(\mathbf{x},t) E_j(\mathbf{x},t)
D_i(\mathbf{x}',0)E_j(\mathbf{x}',0)
\right\rangle_{\mathrm{c}}
=
\left\langle
D_i(\mathbf{x},t)D_i(\mathbf{x}',0)
\right\rangle
\left\langle
E_j(\mathbf{x},t)E_j(\mathbf{x}',0)
\right\rangle
+
\left\langle
D_i(\mathbf{x},t)E_j(\mathbf{x}',0)
\right\rangle
\left\langle
E_j(\mathbf{x},t)D_i(\mathbf{x}',0)
\right\rangle.
\end{equation}
We use the Fourier convention
\begin{equation}
A(\mathbf{x},t)
=
\int_{-\infty}^{\infty}
\frac{d\omega}{2\pi}\,
e^{i\omega t}A(\mathbf{x},\omega),
\end{equation}
and define, due to time-translation symmetry,
\begin{equation}
\left\langle
A(\mathbf{x},\omega)
B(\mathbf{x}',\omega')
\right\rangle
=
2\pi\delta(\omega+\omega')
S^{AB}(\mathbf{x},\mathbf{x}';\omega).
\end{equation}
It follows that
\begin{equation}
\mathcal{I}_{ij}({\bf x}-{\bf x}')
=
\int_{-\infty}^{\infty}
\frac{d\omega}{4\pi}
\Big[
S^{DD}_{ii}(\mathbf{x},\mathbf{x}';\omega)
S^{EE}_{jj}(\mathbf{x},\mathbf{x}';-\omega)
+
S^{DE}_{ij}(\mathbf{x},\mathbf{x}';\omega)
S^{ED}_{ji}(\mathbf{x},\mathbf{x}';-\omega)
\Big],
\end{equation}
where each of these correlation functions has been computed in the preceding section. Using these results we find
\begin{align}
\mathcal{I}_{ij}({\bf x}-{\bf x')}
&=
\frac{T^2}{\pi}
\int_{-\infty}^{\infty}
\frac{d\omega}{\omega^2}
\Bigg[
\Bigg\{
\operatorname{Im}
\left[
\epsilon(\omega)
\epsilon(\omega)
Q_{ii}({\bf x},{\bf x}',\omega)
\right]
-
\operatorname{Im}\epsilon(\omega)
\delta(\mathbf{x}-\mathbf{x}')
\Bigg\}
\operatorname{Im}Q_{jj}({\bf x},{\bf x}',\omega)
\nonumber\\
&\qquad+
\operatorname{Im}
\left[
\epsilon(\omega)
Q_{ij}({\bf x},{\bf x}',\omega)
\right]
\operatorname{Im}
\left[
\epsilon(\omega)
Q_{ji}({\bf x},{\bf x}',\omega)
\right]
\Bigg].
\end{align}

In a homogeneous dielectric medium with frequency-dependent
permittivity \(\epsilon(\omega)\), the  Green's function is simple and satisfies
\begin{equation}
-\epsilon(\omega)\nabla^2
G(\mathbf{x}-\mathbf{x}',\omega)
=
\delta(\mathbf{x}-\mathbf{x}').
\end{equation}
Its Fourier representation is
\begin{equation}
G(\mathbf{y},\omega)
=
\int\frac{d\mathbf{q}}{(2\pi)^3}
\frac{e^{i\mathbf{q}\cdot\mathbf{y}}}
{\epsilon(\omega)q^2},
\qquad
\mathbf{y}=\mathbf{x}-\mathbf{x}'.
\end{equation}
Since
\begin{equation}
\partial_i
=
\frac{\partial}{\partial y_i},
\qquad
\partial_j'
=
-\frac{\partial}{\partial y_j},
\end{equation}
we have
\begin{equation}
\partial_i\partial_j'
G(\mathbf{x}-\mathbf{x}',\omega)
=
-\partial_{i}\partial_{j}
G(\mathbf{y},\omega).
\end{equation}
It follows that
\begin{equation}
Q_{ij}({\bf y})= 
-\partial_i\partial_j
G(\mathbf{y},\omega)
=
\frac{1}{\epsilon(\omega)}\int\frac{d\mathbf{q}}{(2\pi)^3}
\frac{q_iq_j}{q^2}
e^{i\mathbf{q}\cdot\mathbf{y}},
\label{eq:bulk-green-derivative}
\end{equation}
and so $\epsilon(\omega)Q_{ij}({\bf y})$ is real and independent of $\omega$. Consequently
\begin{equation}
\operatorname{Im}
\left[
\epsilon(\omega)Q_{ij}({\bf y})
\right]
=0,
\end{equation}

The time-integrated  stress-stress correlator
\begin{equation}
\Lambda_{ij}(\mathbf{y})=  \int_{0}^\infty dt\left\langle
   \sigma_{ij}({\bf y},0) \sigma_{ij}({\bf 0},t) \right\rangle_c 
\end{equation}
is therefore, using $\operatorname{Im}\epsilon(\omega)
\operatorname{Im}
\left[
\frac{1}{\epsilon(\omega)}\right]= \operatorname{Im}\epsilon^*(\omega)
\operatorname{Im}
\left[
\frac{1}{\epsilon^*(\omega)}\right]= \operatorname{Im}\epsilon(-\omega)
\operatorname{Im}
\left[
\frac{1}{\epsilon(-\omega)}\right]$, given by
\begin{equation}
\Lambda_{ij}(\mathbf{y})
=
\frac{2T^2}{\pi}
\int_0^\infty
\frac{d\omega}{\omega^2}
\operatorname{Im}\epsilon(\omega)
\operatorname{Im}
\left[
\frac{1}{\epsilon(\omega)}
\right]
\left[
-\epsilon(\omega)
\partial_{i}^2G(\mathbf{y},\omega)
-\delta(\mathbf{y})
\right]
\left[
-\epsilon(\omega)
\partial_{j}^2G(\mathbf{y},\omega)
\right],
\label{eq:bulk-stress-kernel}
\end{equation}
where \(i\neq j\). The last two factors are independent of
frequency, since
\begin{equation}
-\epsilon(\omega)
\partial_{i}^2G(\mathbf{y},\omega)
=
\int\frac{d\mathbf{q}}{(2\pi)^3}
\frac{q_i^2}{q^2}
e^{i\mathbf{q}\cdot\mathbf{y}}.
\end{equation}
The viscosity due to dipolar interactions is 
\begin{equation}
\eta_d
=
\frac{1}{T}
\int d\mathbf{y}\,
\Lambda_{ij}(\mathbf{y}).
\end{equation}
Using Parseval's theorem directly in
Eq.~\eqref{eq:bulk-stress-kernel}, we obtain
\begin{equation}
\int d\mathbf{y}\,
\left[
-\epsilon(\omega)\partial_{i}^2G(\mathbf{y})
-\delta(\mathbf{y})
\right]
\left[
-\epsilon(\omega)\partial_{j}^2G(\mathbf{y})
\right]
=
\int\frac{d\mathbf{q}}{(2\pi)^3}
\left(
\frac{q_i^2}{q^2}-1
\right)
\frac{q_j^2}{q^2}.
\label{eq:parseval-bulk}
\end{equation}
For an isotropic cutoff and \(i\neq j\), the angular averaging gives
\begin{equation}
\left\langle
\frac{q_i^2q_j^2}{q^4}
\right\rangle_{\Omega}
=
\frac{1}{15},
\qquad
\left\langle
\frac{q_j^2}{q^2}
\right\rangle_{\Omega}
=
\frac{1}{3}.
\end{equation}
Thus
\begin{equation}
\int\frac{d\mathbf{q}}{(2\pi)^3}
\left(
\frac{q_i^2}{q^2}-1
\right)
\frac{q_j^2}{q^2}
=-\frac{4}{15}
\int\frac{d\mathbf{q}}{(2\pi)^3}.
\end{equation}
Since
\begin{equation}
\operatorname{Im}
\left[
\frac{1}{\epsilon(\omega)}
\right]
=
-
\frac{\operatorname{Im}\epsilon(\omega)}
{|\epsilon(\omega)|^2},
\end{equation}
we obtain
\begin{equation}
\eta_d
=
\frac{8T}{15\pi}
\left[
\int\frac{d\mathbf{q}}{(2\pi)^3}
\right]
\left[
\int_0^\infty
\frac{d\omega}{\omega^2}
\frac{
[\operatorname{Im}\epsilon(\omega)]^2
}{
|\epsilon(\omega)|^2
}
\right].
\label{eq:bulk-viscosity-general}
\end{equation}

We now impose the isotropic Fourier cutoff
\begin{equation}
q<q_{\max},
\qquad
q_{\max}=\frac{2\pi}{a}.
\end{equation}
The momentum integral is
\begin{equation}
\begin{split}
\int_{|\mathbf{q}|<2\pi/a}
\frac{d\mathbf{q}}{(2\pi)^3}
&=
\frac{4\pi}{(2\pi)^3}
\int_0^{2\pi/a}q^2\,dq
\\
&=
\frac{4\pi}{3a^3}.
\end{split}
\end{equation}
Consequently, we arrive at Eq.~(4) of the main text,
\begin{equation}
\eta_d
=
\frac{32T}{45a^3}
\int_0^\infty
\frac{d\omega}{\omega^2}
\frac{
[\operatorname{Im}\epsilon(\omega)]^2
}{
|\epsilon(\omega)|^2
},
\label{eq:bulk-viscosity-cutoff}
\end{equation}
or 
\begin{equation}
\eta_d
=
-\frac{32T}{45a^3}
\int_0^\infty
\frac{d\omega}{\omega^2}
\operatorname{Im}\epsilon(\omega)
\operatorname{Im}
\left[
\frac{1}{\epsilon(\omega)}
\right]
\end{equation}
For a Debye dielectric,
\begin{equation}
\epsilon(\omega)
=
\epsilon_0
+
\frac{\chi}{1+i\omega\tau},
\qquad
\epsilon_s=\epsilon_0+\chi,
\end{equation}
the frequency integral is
\begin{equation}
\int_0^\infty
\frac{d\omega}{\omega^2}
\frac{
[\operatorname{Im}\epsilon(\omega)]^2
}{
|\epsilon(\omega)|^2
}
=
\frac{
\pi\chi^2\tau
}{
2\epsilon_s(\epsilon_s+\epsilon_0),
}.
\end{equation}
This gives Eq.~(5) of the main text,
\begin{equation}
\eta_d
=
\frac{
16\pi T\chi^2\tau
}{
45a^3\epsilon_s(\epsilon_s+\epsilon_0)
}.
\end{equation}
where $\epsilon_s = \epsilon_0 + \chi$ is the static dielectric constant.

\end{document}